# A Benchmark for Image Retrieval using Distributed Systems over the Internet: BIRDS-I

Neil J. Gunther,[a] Giordano Beretta[b]

## Abstract

Comparing the performance of CBIR (Content-Based Image Retrieval) algorithms is difficult. Private data sets are used, so it is controversial to compare CBIR algorithms developed by different researchers. Also, the performance of CBIR algorithms is usually measured on an isolated, well-tuned PC or workstation. In a real-world environment, however, the CBIR algorithms would only constitute a minor component among the many interacting components needed to facilitate a useful CBIR application e.g., Web-based applications on the Internet. The Internet, being a shared medium, dramatically changes many of the usual assumptions about measuring CBIR performance. Any CBIR benchmark should be designed from a networked systems standpoint. Networked system benchmarks have been developed for other applications e.g., text retrieval, and relational database management. These benchmarks typically introduce communication overhead because the real systems they model are distributed applications e.g., an airline reservation system. The most common type of distributed computing architecture uses a client/server model. We present our implementation of a client/server CBIR benchmark called *BIRDS-I* (Benchmark for Image Retrieval using Distributed Systems over the Internet) to measure image retrieval performance over the Internet. The BIRDS-I benchmark has been designed with the trend toward the use of small personalized wireless-internet systems in mind. Web-based CBIR implies the use of heterogeneous image sets and this, in turn, imposes certain constraints on how the images are organized and the type of performance metrics that are applicable. Surprisingly, BIRDS-I only requires controlled human intervention for the compilation of the image collection and none for the generation of ground truth in the measurement of retrieval accuracy. Benchmark image collections need to be evolved incrementally toward the storage of millions of images and that scaleup can only be achieved through the use of computer-aided compilation. Finally, the BIRDS-I scoring metric introduces a tightly optimized image-ranking window, which is important for the future benchmarking of large-scale personalized wireless-internet CBIR systems.

## 1.0 Introduction

At the Internet Imaging Conference 2000 the suggestion was put forward to hold a public contest to assess the merits of various image retrieval algorithms. The contest would be held at the Internet Imaging Conference in January 2001. Since the contest would require a uniform treatment of image retrieval systems, the concept of a benchmark [1–4] quickly enters into this scenario. The contest would exercise one or more such Content Based Image Retrieval (CBIR) benchmarks. The contest itself became known as the *Benchathlon*.

In parallel developments, several authors have also proposed CBIR benchmarks. The most interesting of these, and in some sense the closest to our own work, are the papers by Müller et al. [5] and Leung and Ip [6]. Unlike their proposals, however, our paper presents an *implementation* of a CBIR benchmark together with its attendant methodology. As far as we are aware, our benchmark methodology differs significantly from anything that has been proposed previously. Two distinguishing features mandated by our approach are:

1. Preparation: Manageable compilation and categorization of images to construct a highly *scalable* benchmark collection with potentially millions of images.
2. Runtime: *Scripted control* for ground truth generation, retrieval and response time measurements.

In other words, human intervention is relegated entirely to the preparation phase in our benchmark methodology.

### 1.1 Images vs. Text

Image retrieval can be text based or content based [7]. In text-based image retrieval, the images are annotated and a database management system is used to perform image retrieval on the annotation. This approach has two major drawbacks, the labor required to manually annotate the images, and imprecision in the annotation process. Content-based image retrieval systems overcome these problems by indexing the images according to their visual content, such as color, texture, etc. Since it is not known how the human visual system indexes by content there is no accurate representation, only estimates. A goal in CBIR research is to design representations that correlate well with the human visual system. Bench-marks facilitate the comparison of various visual representations and they also provide other relevant metrics e.g., retrieval response time. CBIR benchmarking based on these representations re-introduces the indexing problems of text-based retrieval in the data preparation phase of the benchmark. BIRDS-I does not solve this problem but it does isolate the issue so that we can address it through the use of ontologies [8], described in Section 2.4 on page 7 of this report.

## 1.2 System Performance

In the case of non-distributed image retrieval systems the usual engineering focus is on tuning each subsystem under the assumption that the system as a whole perform better. However, this will not be true in general because the narrow perspective of tuning a subsystem cannot take into account the possible interactions of all the subsystems. Therefore, it is important to optimize the system as a whole. This global or holistic approach can have the counter-intuitive consequence that the subsystem performance should actually be tuned sub-optimally [9]. Consider your own PC (Personal Computer) and the constant barrage of advertising to upgrade the processor to the fastest speed available to improve performance. But the CPU (Central Processing Unit) is only one component in the PC system. Unless your application is very CPU-intensive (e.g., image rendering), you can end up suffering the classic syndrome of *hurry up and wait*. And it should be kept in mind that all applications wait at the same speed! The faster processor may be forced to wait for the now relatively slower retrieval of an image from the PC disk system. For image editing applications it is often more prudent to upgrade either the physical memory capacity or the disk speed rather than upgrading the processor.

For more complex computer systems like multi-user timeshare systems, multi-processor servers, and multi-server clusters, the interactions are also more complex [10]. Compared to the single-threaded PC (whose behavior is generally sequential and deterministic) larger scale systems are often multi-threaded to support parallel or concurrent processing and are therefore non-deterministic. This makes system tuning less predictable and leads naturally to empirical evaluation using standardized benchmarks [11, 12] of the type that have been used for assessing large-scale systems.

Benchmarking typically involves measuring the performance of the system as a whole. A benchmark like BIRDS-I treats the computer system as a black box [3]. The major unknown in the Benchathlon system is the CBIR application, so the contest platform is therefore only tuned in a generic way. Each contestant strives to optimize their own CBIR application prior to running it on the Benchathlon platform. The BIRDS-I benchmark measures the performance of the CBIR application by executing the predefined benchmark queries and then reporting summary metrics such as retrieval accuracy scores and query response times. The same exercise is repeated for each contestant. The benchmark runs can then be used to rank the performance of the CBIR application belonging to each contestant.

Benchmarks not only allow us to compare different systems, and evaluate the solution best suited for a particular problem, they also allow system architects to identify performance bottlenecks and design systems that scale well. On the Internet it is particularly difficult to predict server utilization. Web site scalability is a perennially unpredictable problem [13].

### 1.3 Layout of This Paper

The organization of this paper is as follows. In Section 2.0, "Benchmark Architecture," on page 5 we present the motivation for the benchmark architecture discussed subsequently. The BIRDS-I benchmark is designed with the trend toward the use of small personalized wireless-networked systems clearly in view [14]. Personalization carries with it many attributes but key among them is the use of images and various forms of CBIR. That personalized mobile computing devices are both small and networked presents challenges for the way CBIR can be used by them. Moreover, the kind of images one finds on the Web tend to be very heterogeneous collections and this imposes certain constraints on how these image should be organized and the type of retrieval accuracy measures that can be applied to gauge CBIR performance.

In Section 3.0, "Benchmark Run Rules," on page 8 we describe the rules for organizing such heterogeneous image collections into versioned files with a file system directory structure. An important consequence of this approach is that BIRDS-I only requires controlled human intervention for the compilation of the image collection and none for the generation of ground truth used in the measurement of retrieval accuracy. Any human intervention needs to be computer-aided and controlled in order to manage the latent complexity of large-scale CBIR systems. At runtime, CBIR benchmarking collections that contain tens of thousands images cannot be interrupted or obscured by the subjective opinions of human judgement. Moreover, benchmark image collections need to evolve incrementally toward millions of images. Similar collection sizes already exist for text retrieval benchmarks [15] and are expected to appear for image repositories on the Web. Incremental collection scaleup to this order can only be achieved through the use of computer-aided compilation.

Section 4.0, "Systems Perspective," on page 9 enumerates the highly distributed computing architectures employed on the Web and Internet. The most ubiquitous of these is known as, *client/server* [16]. Personalized wireless computing (recognized in Section 2.0 on page 5) is the latest example of a client/server application. Section 4.0 presents taxonomy of client/server architectures that can be expected to arise in future CBIR benchmarks. We also identify where BIRDS-I falls in this taxonomy.

Section 5.0, "Benchmark Metrics," on page 13 presents a detailed discussion of the retrieval-scoring scheme used in BIRDS-I. Two concepts we introduce into the scoring metric are a tightly optimized image-ranking window, and a penalty for missed images. Some numerical comparisons with other scoring schemes are provided. We believe each of these considerations is important for the automated benchmarking of large-scale CBIR.

## 2.0 Benchmark Architecture

### 2.1 Application Context

Traditional image retrieval system design has implicitly relied in the availability of cheap processing power. Moore's law [9], which states that CPU processing power doubles every eighteen months, supports this assumption. This design approach is further epitomized by video-game machines, which for a very modest price offer specialized graphical processing power far superior to that available in a typical PC. The capacity of disk drives doubles every year and the speed of I/O connections is also rapidly increasing. Finally, Internet access is becoming faster and cheaper [17]. These numbers raise the user's expectations for a satisfactory user experience.

Contrast this, however, with the trend toward small mobile Internet devices (such as palm-tops and personal digital assistants), where the communications channels have lower bandwidth and the devices themselves have only small screen real-estate [17]. Combined with their low power requirements and low purchase price, these mobile devices will allow a larger portion of the world's population to use the Internet in an even more personalized way than has been true heretofore with the PC. Multimedia in general and images in particular, are expected to play a significant role in defining such personalization.

From these observations about a mobile wireless world, we conclude that it will become less acceptable for users to view a large number of retrieved images and intolerable for them to be subjected to narrowly scoped queries that return a large number of incorrect images. This conclusion ramifies into our benchmark methodology and the scoring scheme we present in Section 5.0 on page 13 and runs counter to some other CBIR benchmark proposals [3].

### 2.2 Scope

Within imaging sciences, such as CBIR, the primary interest is in images encoded as signals. This stands in contrast to the approach taken in the vision sciences where object recognition is the primary interest [18]. In other words, in CBIR we are interested in the signal and not its semantics. This difference narrows the scope of what our benchmark needs to do.

Furthermore, we are not interested in text-based image retrieval, because this case is covered by benchmarks for conventional text retrieval systems [11, 19]. We assume that no metadata is stored explicitly in the image collection and only content-based image retrieval algorithms will be applied. A peculiarity of CBIR algorithms is that queries can return a similarity set instead of an exact match. For a comprehensive treatment of state-of-the-art image retrieval we refer the reader to the excellent review by Rui, Huang, and Chang [7].

On the Internet, the interesting image retrieval applications are those implemented on the World Wide Web [20–23]. In this context, the nature of the images stored in collections is unknown a priori. We are therefore interested in heterogeneous image collections rather than domain-specific collections. The implicit assumption is that we are interested in algorithms indexing general visual features (e.g., color, texture, shape, faces, etc.). Algorithms that index domain-specific features can be expected to have poorer performance because the overhead in matching domain-specific features is not advantageous for a general image collection.

The constraints imposed on the scope of the benchmark described above must now be reflected in the way the images in the collection are grouped into similarity sets.

### 2.3 Image Collection Structure and Ground Truth

As mentioned in the introduction, scalability is an important issue for Internet applications and a motivation for benchmarking during the application design phase. A fortiori, any Internet-based CBIR benchmark must be extensible in order to measure the scalability of the CBIR application under test. This implies the initial benchmark should employ a large image collection combined with the ability to grow that test collection over time [20]; a requirement long understood in the text retrieval community [15].

We propose to start with an image collection of at least 10,000 images[c] so as to prohibit them from being entirely memory resident or otherwise cached on the platform or the Internet. Starting with such a large number of images, together with the need to unambiguously assess the retrieval precision and recall with reasonable efficiency, demands a careful organization of the image collection.

Image collections based on stock photographs consist of normalized images that have been photographed under controlled conditions with possible additional processing. In general, we cannot expect this to be the case on the Internet, where images typically suffer from artifacts such as incorrect tone reproduction and under-correction for chromatic adaptation [8]. Therefore, it is imperative that the benchmark image collection consist of an eclectic and heterogeneous image compilation.

Precision and recall are determined with the aid of a data structure known as the ground truth, which embodies the relevance judgments for queries [5]. For exact matches the ground truth is easy to establish: the file identifier of the relevant image is the same as the identifier of the query image. For similarity matches the ground truth is harder to define because it is necessary to categorize the image collection into sets of similar images. Therefore, two problems must be solved.

---

c. This number is based on intuition. When in the early 1980s the first photographic cameras with matrix exposure methods came on the market, databases of 10,000 images were used to recognize the scene illumination type to determine the optimal exposure parameters. With the availability of inexpensive RAM chips, manufacturers over time increased the size ten-fold, but lately the number has decreased to around 30,000 images as better algorithms have been invented. This application deals just with illumination conditions; a CBIR system has many more parameters and the database must contain a much larger number of images. 10'000 is an achievable starting point.

First, appropriate image categories must be defined. Second, the images themselves must actually be categorized according to those definitions.

### 2.4 Categorization Problem

Regarding the first problem, it is easy to define a small and finite set of categories with a theme such as, a naïve enumeration of life's progression viz.: births, christenings, birthdays, graduations, weddings, and funerals. However, a typical family may actually have a few thousand photos to categorize over a lifetime. If the family were to try and use this small set of categories, they would quickly discover photos that did not fit into any of these categories e.g., photos of the family car.

Jörgensen [24] achieved a major breakthrough by introducing a method based on Gestalt principles of organization[d] for constructing a set of images categories. In essence, a number of people are given the task of iteratively refining the categorization of a set of images as a controlled experiment until they can easily retrieve the images again. The resulting categories derived from all participants are finally pooled so that a consensus categorization can be identified. However, there is scaling limitation to this approach. Whereas it facilitates the creation of meaningful categories, when the number of categories gets large (just a few hundred) it becomes difficult for humans to locate the category itself.

To overcome this limitation, a partial ordering could be imposed on the categories to develop a hierarchy. However, in the case of images there is no known natural order so it becomes difficult to define a comprehensive hierarchy. Every category has to be referenced with respect to all other categories.

Instead of trying to construct a single taxonomy of all possible image categories (i.e., introducing a single relation between all categories), we can partition the sorting task by a number of independent taxonomies or relations (e.g., life events, family members, family objects, etc.) to produce an ontology [25, 26]. In the family photo problem mentioned above, the family car is now more easily identifiable as belonging to the smaller, more manageable taxonomy of family objects.

Tillinghast and Beretta [8] used this approach to categorize images to provide a scalable method that allows users to create and maintain extensive systems of categories. This methodology scales particularly well when compared with taxonomies because it is possible to define small local structures and stitch them together. This ontology approach is becoming more prevalent in Web services e.g., Yahoo.com, and AskJeeves.com.

### 2.5 Directory Structure

As for the second problem, we now lay out the rules for defining the ground truth in the Benchathlon contest. We have to be particularly careful that it be possible to repeat a benchmark at some time in the future, e.g., with an improved version of a CBIR algorithm.

---

d. A conception of organization based on the idea that a perceived whole is not just the sum of its parts but rather the parts are organized into structures in principled ways.

We create a file structure where there is a subdirectory for each image category, similar to that being done for the collection under way at the University of Washington in Seattle [27]. Each image is categorized by moving it into the appropriate subdirectory. An image can be copied into more than one subdirectory if it can be attributed to more than one image category according to its visual content. The subdirectories can be nested if there is a strong reason for not introducing a separate category, but this should be the exception, because structure and contents should be independent.[e]

A malicious participant could exploit this disk file structure to cheat in the benchmark [1]. To avoid this problem a separate query directory is created and used for the actual queries. Each entity in this directory is a link to an image in the collection. The link identifier is computed with a fixed hash algorithm so that a given image will always have the same link identifier.

Once a categorization process has been completed, a script is executed to enumerate the subdirectories, generate the query directory contents, and create a versioned file containing a list of the images in each category. This versioned file is the ground truth, independent of the contents of the file structure.

## 3.0 Benchmark Run Rules

Categorization should be carried out by trained domain experts who, like the physicians in the example cited by Müller et al. [5], are specially trained in this task and provide an authoritative verdict. The training is crucial for the value of the image collection, because once an image has been categorized, it cannot be categorized again. We saw in the previous section how the ground truth is described in a versioned file. Versioning allows for appending images to the collection without affecting the benchmark results over time. If an image were removed from a category in a reclassification, all ground truth files would become invalid and experiments could not be repeated.

The categorization training has to be very specific in teaching visual similarity and how it differs from semantic, semiotic, and other ontological equivalence classes. The training must be as serious as that of a pathologist who must be able to identify a lump in a mammogram, because the viability and value of the Benchathlon and its benchmarks depend on it.

The categorization is conservative. When classifiers have a doubt, they should seek the opinion of one or more of the Benchathlon organizers. If the category for a particular image is not immediately clear to an organizer, then the image should not be included in the image collection.

The benchmark is run by executing scripts. No human is involved at query time thus, the process is completely automatic. Aspects of CBIR applications such as the readability of the result or the user interface of the application are not the sub-

---

e. A possible reason for creating subdirectories could be to balance the directory structure; this could improve the performance of the system on which the benchmark is executed.

ject of this benchmark; there are other methodologies, such as usability testing, to cover these user experience metrics.

All files, including the image collection, the versioned ground truth files, and the scripts for each benchmark must be freely accessible on the World Wide Web. For this purpose we have registered two domain names: benchathlon.net is for the Web server containing everything that is necessary to perform a Benchathlon event, benchathlon.org is for a Web server to be used for the Benchathlon organizers to coordinate their work.

An attribute of the personalized Web experience is near-instantaneous response, including the retrieval of images. Therefore, any CBIR benchmark must provide metrics for response time performance. (See [28] for a more extensive discussion).

## 4.0 Systems Perspective

The traditional focus of CBIR performance comparisons has been on the image retrieval algorithms themselves. These algorithms run on either a stand-alone platform such as a PC or workstation. The typical performance constraints are narrowly confined to just the CPU speed and memory capacity (and possibly the memory access speed). The advent of Web-based image retrieval systems [20–22] suggests this focus on algorithmic performance has become too narrow [5, 6].

Our philosophy is captured in the choice of name for the first implementation of such a benchmark viz., BIRDS-I an acronym for Benchmark for Image Retrieval using Distributed Systems over the Internet. This name is intended to invoke the dual notions of:

1. A high altitude systems view encompassing the interaction of various CBIR subsystem components
2. A high degree of acuity to resolve the performance differences between CBIR algorithms

Hence the choice of an eagle's eye for the web page logo (http://benchath.hpl.external.hp.com/).

The BIRDS-I development system represents one of the simplest examples of a client/server system viz., a client platform connected to remote services across a network (in this case the Internet). The retrieval algorithms and the data collection are treated as a black box subject to measurement by the benchmark. Benchmark metrics include measurements of accuracy and response time [6]. Another goal of the BIRDS-I benchmark design was to make it self-contained without the need for human subjective intervention with regard to measuring retrieval accuracy.

Our approach of measuring client/server networked architectures [16, 29] follows directly from our philosophical conviction that client/server is key to the future of Internet applications (especially wireless/mobile). Therefore, benchmark-ing CBIR and all aspects of that architecture will need to be included in any future CBIR benchmark designs.

### 4.1 Client/Server Systems Taxonomy

Because of our emphasis on measuring the performance of the system architecture in the BIRDS-I benchmark, it will aid the subsequent discussion to give a brief taxonomy of client/server architectures. The following taxonomy is also intended as a kind of architectural roadmap for future CBIR benchmark designs.

From a computer science standpoint, "client/server" is a formal *software* interface definition [9] in that a client makes a procedure call to a set of defined operations or methods that support a service. The most familiar example of a client/server operation is a Web browser making HTTP GET requests to an HTTPd server that listens for such requests.

What is often overlooked, however, is the fact that the browser client and the HTTPd server can be running on the *same* platform and communicating via the HTTP protocol without the network packets actually traversing any physical network.

A client/server interface does not define where the services must run, nor anything about how remote services should be networked. The network could use an ethernet protocol or an internet protocol or both. Having said that, there is no doubt that the most ubiquitous and useful variant of client/server architectures does employ remote services over a physical packet network. Once again, the Web is the most obvious example of this generic client/server architecture in Figure 1.

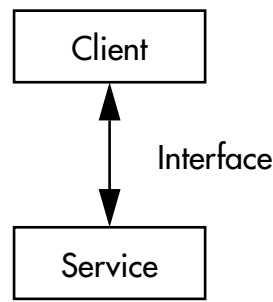

**FIGURE 1.** *Schematic of a canonical client/server architecture.*

#### 4.1.1 Single User/Client

A further useful step in system organization is to differentiate between software processes and the hardware resources they run on.

Figure 2 shows a single user process running on PC hardware and accessing a database process on a remote server connected via a physical network. This is typical of most Web browser setups. It also happens to be the BIRDS-I development environment. The client process is the BIRDS-I benchmark script running under Perl, the network is the Internet and the services provided by the Benchathlon CBIR server at HP was 9 hops away on the Internet (on average) from the client platform.

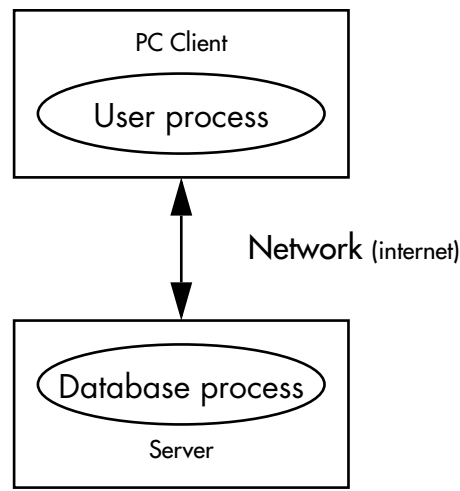

**FIGURE 2.**   *Schematic of user and database precesses.*

It is important to note from a benchmarking standpoint that this client/server configuration does not impinge on system scalability issues [9, 30]. There is no attempt to represent more the one user accessing the database. Hence, there is possibility of testing concurrency constraints, serialization of database accesses or other potential system bottlenecks [1, 9].

In the current implementation of the BIRDS-I benchmark, only queries are issued. There are no database transactions that concurrently update the database (e.g., adding new images to the data collection). Queries are read-only, so they can be cached. Queries can also be processed in parallel [10] assuming that image collections can be partitioned in an appropriate way. Cf: TerraServer, Corbis.com, ditto.com, etc.

Moreover, the data that is retrieved corresponds to image names, not the image file content. Image files are not transported across network [16] nor rendered on the client-side process. In this sense, the issue of how to best assess the quality of the type of image returned is does not arise in the BIRDS-I benchmark. Neither BIRDS-I, nor the Benchathlon contest, is currently expected to address these issues but they are fully expected to be part of future developments.

#### 4.1.2 Multi-User/Client

In order to test scalability issues, more user processes need to be added to the client side. A system like that shown schematically in Figure 3 on page 12, repre-sents a more complex CBIR system where geographically dispersed users can access the same image collection.

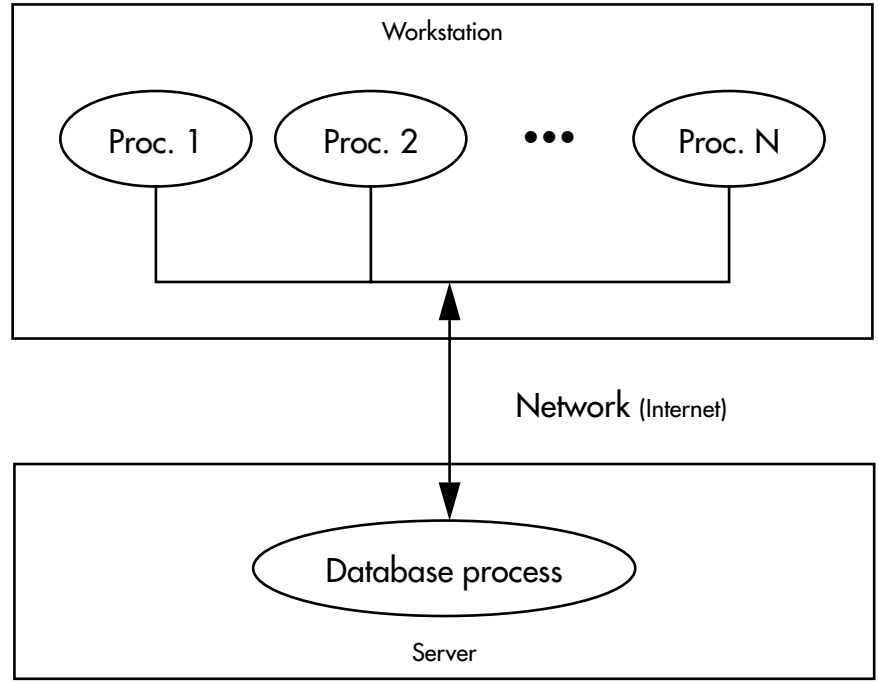

**FIGURE 3.**  *Multi-user client/server architecture.*

System driver constraints are determined by such components as the client-side CPU, memory, and network interfaces. Server-side constraints arise from the image database retrieval system's ability to serve multiple requests e.g., if the database is single-threaded as well as how many queries or transactions can be handled in parallel.

### 4.1.3 Multi-User/Multi-Client

Finally, client-side user processes can be scaled up further by partitioning them across multiple client platforms, and network interfaces as depicted in Figure 4. This is the kind of Benchathlon system that would mimic global-scale Web-based CBIR.

Here, scalability is mostly constrained by server-side database system's ability to support multiple network processes and respond to multiple users.

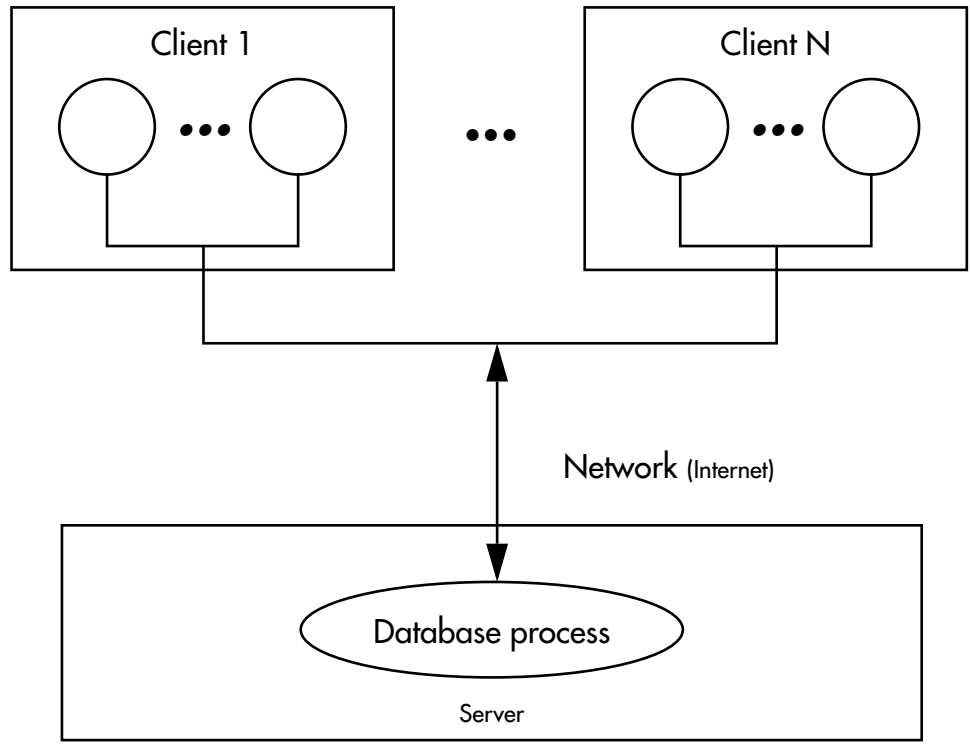

**FIGURE 4.**  *Multi-user multi-client architecture.*

## 5.0 Benchmark Metrics

The BIRDS-I benchmark measures both CBIR accuracy and query response time. Several authors have proposed retrieval metrics based on well-known text retrieval Precision-Recall (PR) graphs. Leung and Ip [6], for example, have proposed using incrementally evaluated PR measures for a CBIR benchmark that involves human assessments staged between queries. The BIRDS-I run rules (see Section 3.0 on page 8) eliminate the need for any human intervention in the retrieval measurement process.

Müller et al. [5] have proposed the use of multiple metrics that include PR graphs together with a ranking metric normalized against the ground truth images. We concur with the use of such a broad set of metrics to provide a more complete performance picture. BIRDS-I uses a normalized rank similar to that proposed by Müller et al. [5] but with the additional feature of penalizing missed images. We view this metric as a leading indicator for making first-order comparisons between CBIR systems. We expect BIRDS-I to evolve toward the use of multiple metrics in the future.

A single leading performance indicator should be provided with any system benchmark otherwise, a less reliable ad hoc metric simply will get invented for making first-order comparisons between CBIR benchmark results. In the case of the SPEC benchmark [12] the SPECint or SPECfp aggregate performance metric is the leading indicator used for making first-order comparisons between platforms but more detailed performance metrics are available for more detailed scrutiny, if necessary. In the case of the TPC-C benchmark [11], the "performance pair" comprising the transaction-throughput and the price-per-transaction are the leading comparison metrics but complete benchmark disclosure reports (including performance graphs) are available from TPC. The normalized ranking (defined in the next section) appears to be a good candidate for such a leading performance indicator in CBIR benchmarking.

### 5.1 Retrieval Scoring

Let there be $q = 1, 2, \ldots, Q$ queries made against an image database. In addition, let $G(q)$ be the number images contained in a vector of ground truth images associated with query $q$ and further denote by $G_{max}$ the largest of these ground truth image vectors. Invoking a query against the total image collection causes an arbitrary set of images to be returned to the client. In the best case, for a query ($q$), all the images belonging to that ground truth vector $G(q)$ would be returned as an exact match of the ground truth vector sequence and would thus correspond to a perfect retrieval score.

However, most image retrieval algorithms are less than perfect, so images that are members of $G(q)$ may be returned:

1. out of order
2. in correct sequence but interspersed with incorrect images
3. as an incomplete subset when not all the images of $G(q)$ are found
4. not at all (worst case)

We need to account for such possibilities in any scoring metric. The basis of our scoring scheme is a ranking procedure. A scoring window $W(q) > G(q)$ is associated with the query ($q$) such that the returned images contained in $W(q)$ are ranked according to an index $r = 1, 2, \ldots, W$ depicted in Figure 5.

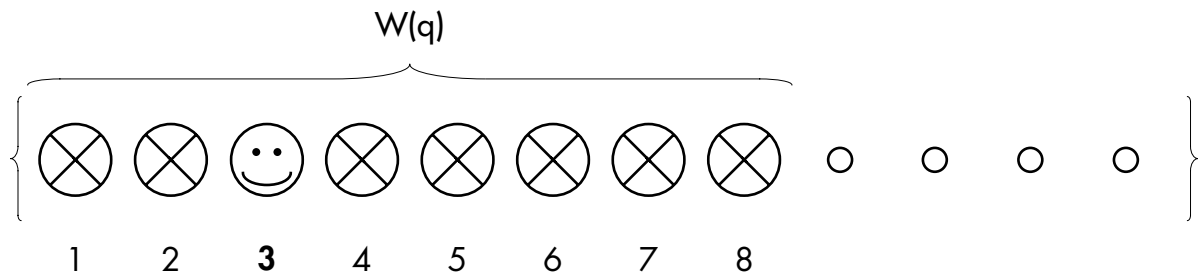

**FIGURE 5.**   *Returned image set with scoring window W and one correct image of rank 3.*

We introduce the step function,

$$\theta(w-g) = \begin{cases} 1 & \text{iff: } w \sim g \\ 0 & \text{otherwise} \end{cases} \quad \text{(EQ 1)}$$

which is zero unless there is a "match" (denoted by '~') between the retrieved image at index *w* and any ground truth image *g*.

The number of *correct* images returned in the window *W(q)* is given by,

$$F(q) = \sum_{w=1}^{W} \theta(w-g) \quad \text{(EQ 2)}$$

and the number of *missed* images returned is,

$$\mu(q) = G(q) - F(q) \quad \text{(EQ 3)}$$

Now, the retrieval rank *R(q)* can be defined as,

$$R(q) = \sigma(q) + \mu(q) \cdot \pi(q) \quad \text{(EQ 4)}$$

where

$$\sigma(q) = \sum_{w=1}^{W} w \cdot \theta(w-g) \quad \text{(EQ 5)}$$

is the sum of the ranks of the correct images and $\pi(W)$ is a penalty for the missed images. Since missed images lie outside the scoring *W(q)*, the value of the penalty must exceed the rank of the last entry in *W(q)*. The choice of penalty weight is arbitrary. For example, Salembier and Manjunath [31], and Manjunath, et al. [32] choose $\pi(W) = 1.25 \cdot W(q)$. In the subsequent discussion, we shall use $\pi(W) = W(q) + 1$.

It is important to note that the value of the retrieval rank *R(q)* is affected only by the position of correct images in the scoring window *W(q)* not their order with respect to the sequence specified by the ground truth vector. If *A* and *B* are correct images in the ground truth set, then the retrieved sets {*A, B*} and {*B, A*} have equal retrieval rank.

A relative retrieval rank *RR(q)* can be defined as the ratio of the retrieval rank with respect to the number of ground truth images in the vector *G(q)* viz,

$$RR(q) = R(q)/G(q). \quad \text{(EQ 6)}$$

In the case of a perfect score, the number of images found by the query *F(q)* equals the number of images in the ground truth vector, i.e., *F(q) = G(q)*, hence, $\mu(r) = 0$ and the retrieval rank is the sum of the successive rank values viz., $G(G + 1)/2$. Substituting this value into the relative rank equation, we find for the *best* relative retrieval rank is given by,

$$RR_b(q) = \frac{1 + G(q)}{2}. \quad \text{(EQ 7)}$$

In the worst case, no ground truth images are returned in the window $W(q)$ so the number of incorrect retrievals is $G(q)$. The retrieval rank $R(q)$ becomes the product $G(q) \cdot \pi(W)$. The *worst* relative retrieval rank is then given by,

$$RR_W(q) = 1 + W(q). \quad \text{(EQ 8)}$$

These extremes in rankings define an interval, $RR_W(q) - RR_b(q)$, within which any relative retrieval rank $RR(q)$ must lie. For the purpose of comparisons, it is preferable to normalize this interval onto the unit interval $NRR(q) \in [0...1]$ via the ratio,

$$NNR(q) = \frac{RR(q)}{RR_W(q) - RR_b(q)} = \frac{RR(q)}{1 + W(q) - \frac{1}{2} \cdot (1 + G(g))} \quad \text{(EQ 9)}$$

If all the ground truth images are returned, $NRR(q) = 0$ (i.e., a perfect score with no penalties) while $NRR(q) = 1$ corresponds to the maximum penalty when no correct images are returned. The normalized average ranking over all queries in the ground truth file produces the retrieval score,

$$S = \frac{\sum_{q=1}^{Q} NNR(q)}{Q}, \quad \text{(EQ 10)}$$

which also lies in the range: $0 \leq S \leq 1$. This is the scoring metric used in the BIRDS-I benchmark.

Since the number of images returned is unknown, the range of retrievals for different queries can vary enormously. This variance can be reduced by selecting a search window of size $W(q)$ proportioned to each query's ground truth vector $G(q)$. The search window paradigm has been used to construct a normalized ranking for a retrieval scoring metric in the context of certain MPEG–7 experiments on color representations [31, 32]. In that scheme, a perfect score gets zero, while a score of 1 means none of the ground truth images are returned.

We adopt a similar approach to scoring retrievals in the BIRDS-I benchmark but differ significantly from other metrics in the way the search window $W(q)$ is chosen. The retrieval score is a very sensitive issue in the context of a public benchmarking and in real-world image retrieval where the user device is small and mobile (like a personal digital assistant). For this reason, we compute a carefully *optimized W(q)* that produces a more realistic scoring procedure.

## 5.2 Optimization Problem

Determining the size for *W*(*q*) is an optimization problem where the goal is to choose a window such that it:

1. minimizes the difference in length between itself and its respective ground truth vector *G*(*q*).
2. minimizes the size of all *W*'s across all queries (*q*).
3. satisfies a positivity condition: [*W*(*q*) – *G*(*q*)] > 0. *W* cannot be smaller than *G*.
4. should be large if *G*(*q*) small, and small if *G*(*q*) is large.

Although 3 and 4 are corollaries to objectives 1 and 2, it will facilitate the subsequent discussion to enumerate them collectively. The difficulty of meeting these objectives simultaneously can be appreciated better if we first eliminate some naïve but incorrect solutions.

### 5.2.1 Naïve Solutions

**Naïve solution (a).** We could meet the requirements 1 through 4 by simply taking the arithmetic mean of all the *G*(*q*) and setting $W = G_{mean}$. This solution is, however, unphysical because it violates the positivity requirement 3 i.e., it's nonsense to choose a search window that is *smaller* than the number of ground truth images. Therefore, this solution can be completely dismissed.

**Naïve solution (b).** Setting *W*(*q*) = *G*(*q*) satisfies objectives 1 and 2 but, once again, violates the positivity condition 3. We require *W*(*q*) to be *larger* than its corresponding *G*(*q*) because the correct images are most likely to be interspersed with incorrect returns.

**Naïve solution (c).** Setting the search window to be a simple multiple of the largest ground truth vector e.g., $W = 2G_{max}$ would make all the windows of equal length and thereby minimize the difference in size between the *W*'s as specified in objective 2. It also satisfies goals 3 and 4 but fails completely to meet objective 1.

In summary, we seek a solution that lies between the extremes represented by naïve solutions (b) and (c).

### 5.2.2 Plausible Solutions

The scoring window that has been used by MPEG–7 researchers [32] for various color and texture experiments is defined as $W_{mpeg}(q) = Min(4 \cdot G(q), 2 \cdot G_{max})$ and is shown graphically in Figure 6. Clearly, it conforms to corollary 4 crudely. Moreover, $W_{mpeg}(q)$ has a discontinuity at $G = G_{max} / 2$ and thus fails to satisfy requirement 2.

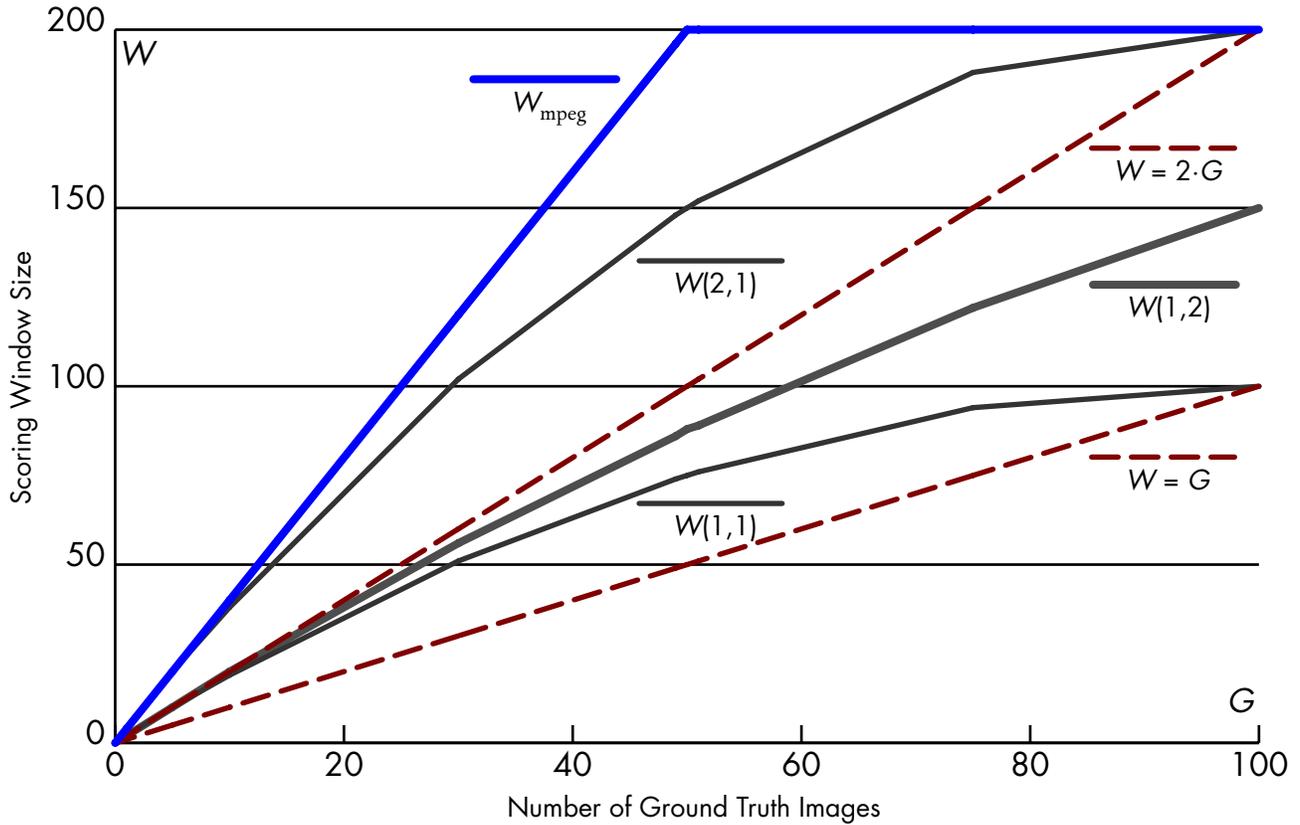

**FIGURE 6.**  *Retrieval window functions. The MPEG–7 function is labeled $W_{mpeg}$.*

On further reflection, we see that objective 4 suggests that we should look for a continuous (or piecewise continuous) *convex* function from which we can select the values for $W(q)$. We propose a class of such convex functions (Figure 7) having the general form:

$$\frac{W(q;k,m)}{k} = \frac{m \cdot G_{max} - (G(q) - (m \cdot G_{max}))^2}{m \cdot G_{max}}; \forall k, m = 1, 2 \qquad \text{(EQ 11)}$$

where the integer-valued search window is chosen as $W(q) = \text{Ceil}(W(q; k, m))$. The mathematical derivation and proof will be given elsewhere.

A benign assumption used in the choice of these functions is that we are only interested in functions that involve simple multiples of $G_{max}$. More elaborate criteria could be established but with seemingly little significance for the problem at hand.

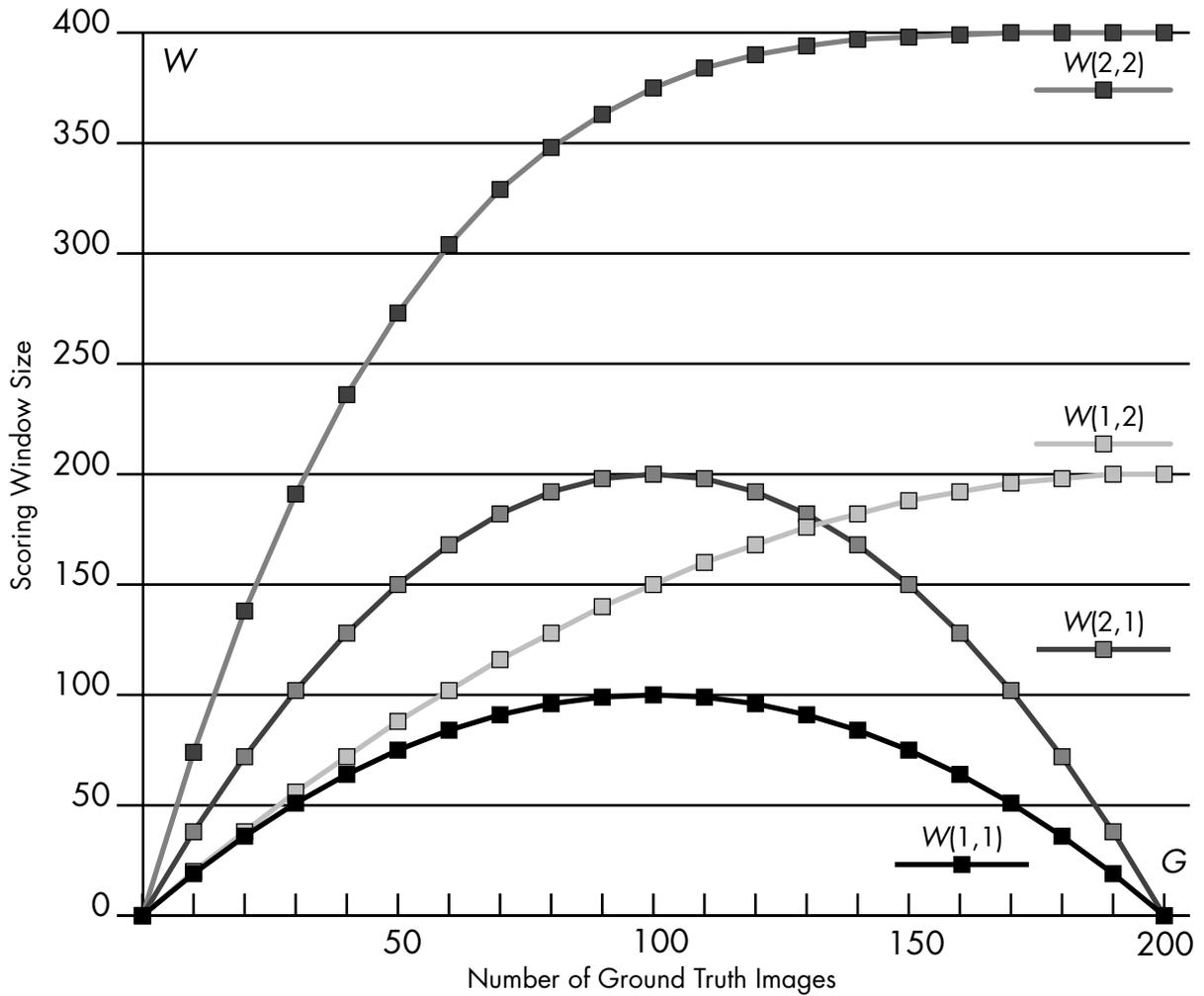

**FIGURE 7.**  *Generalized class of convex scoring functions.*

As can be seen in Figure 7, this class of convex solutions corresponds to a family of inverted parabolic functions. They meet the positivity condition: $[W(q) - G(q)] > 0$, and corollary 4 that the $W(q)$ should be large for small $G(q)$, and vice versa. More particularly, a parabolic function has the unique feature that the change in slope (the measure of the successive difference in lengths of the *W*'s) is constant and therefore minimizes those differences across all *W*'s as required by objective 2. The only remaining issue, therefore, is how best to meet minimization objective 1. We do this by examining each of the discrete cases corresponding to the indices *k*, *m* taking on the values 1 and 2 successively.

### 5.3 Numerical Comparisons

The reader should refer to Table 1 with $G_{max} = 100$ for the following discussion.

1. Although piecewise convex, *W*(*q*: 2,2) lies well above the range of the $W_{mpeg}(q)$ function in Figure 7 and therefore we do not examine its properties any further.
2. *W*(*q*: 2,1) is an improvement over the $W_{mpeg}(q)$ function in that it is piecewise convex and contains no discontinuities (see Table 1). However, it can be demonstrated that it does not offer the tightest bound across all queries *Q*.
3. *W*(*q*: 1,1) meets objectives 1 and 2 but it suffers the drawback that it violates the positivity condition at one point; the endpoint $G = G_{max}$. In other words, for the query with the largest ground truth vector, it is too close for comfort!
4. *W*(*q*: 1,2), on the other hand, satisfies all of the conditions 1 through 4 for all *G*(*q*) in [0, $G_{max}$].

| G | $W_{mpeg}$ | W = G | W = 2·G | W(1,1) | W(1,2) | W(2,1) |
|---|---|---|---|---|---|---|
| 0 | 0 | 0 | 0 | 0 | 0 | 0 |
| 1 | 4 | 1 | 2 | 2 | 2 | 4 |
| 5 | 20 | 5 | 10 | 10 | 10 | 20 |
| 10 | 40 | 10 | 20 | 19 | 20 | 38 |
| 30 | 120 | 30 | 60 | 51 | 56 | 102 |
| 49 | 196 | 49 | 98 | 74 | 86 | 148 |
| 50 | 200 | 50 | 100 | 75 | 88 | 150 |
| 51 | 200 | 51 | 102 | 76 | 89 | 152 |
| 75 | 200 | 75 | 150 | 94 | 122 | 188 |
| 100 | 200 | 100 | 200 | 100 | 150 | 200 |

**TABLE 1.** *Comparison of image retrieval scoring windows.*

In every case, it is quite clear that the optimized function *W*(*q*: 1, 2) offers a tighter scoring window than $W_{mpeg}(q)$ and the tightest *W*(*q*) for all relevant integral multiples of $G_{max}$.

Based on our conclusions about mobile wireless-internet CBIR described in Section 2.0 on page 5, it will become less acceptable for users to view a large number of retrieved images and intolerable for them to be subjected to narrowly scoped queries that return a large number of incorrect images. The optimized search window defined in this section satisfies that requirement.

## 6.0 Summary

The BIRDS-I benchmark has been designed with the trend toward the use of small, personalized, wireless-networked systems clearly in mind. Personalization with respect to the Web implies heterogeneous collections of images and this, in turn, dictates certain constraints on how these images can be organized and the type of retrieval accuracy measures that can be applied to CBIR performance.

BIRDS-I only requires controlled human intervention for the compilation of the image collection and none for the generation of ground truth for the assessment of retrieval accuracy. Benchmark image collections need to be evolved incrementally toward millions of images and that scaleup can only be achieved through the use of computer-aided compilation. Increasingly, the most ubiquitous distributed computing architectures employed on the Web are client/server, personalized wireless computing being a foremost example. Two concepts we introduced into the scoring metric are a tightly optimized image-ranking window, and a penalty for missed images, each of which is important for the automated benchmarking of large-scale CBIR.

The Perl scripting language [33] was used to expedite network programming [34] of the client/server-based image retrieval system. The client platform was an Apple Graphite iMac® personal computer running the MacPerl 5.3.5 development environment [35]. Although MacPerl 5 is a single-threaded application in that only a single user process maybe running, it was sufficient to expedite the development of the prototype BIRDS-I benchmark script. So far, we have only implemented a simple client query version of the client/server architecture without invoking image file transport or image rendering.

On the server side, a set of over 3,700 unencumbered heterogeneous images from a previous project [8] was made freely available at http://www.hpl.hp.com/personal/Giordano_Beretta/Albums/rolls.html together smaller sets of images contributed by other interested parties. A the time of this writing, these images have not yet been categorized for visual similarity.

The modus operandi for the Benchathlon is open participation. Interested parties communicated via a public list server that can be accessed at http://www.topica.com/lists/benchathlon/read. The server for the Benchathlon is also publicly available, in the sense that an e-mail message to beretta@hpl.hp.com or ullas@hplug.hpl.hp.com is all that is required to get an account on http://benchath.hpl.external.hp.com. We currently have also 20,000 Corel images and 13,000 other color images. Sample query images and associated ground-truth (the correct answers) are also available and more will be generated for scaled/rotated and cropped image query types by simple image manipulation.

## 7.0 Acknowledgements

The authors are indebted to the following colleagues for their contributions to this work. The original suggestion for holding an image retrieval contest was due to Theo Gevers (University of Amsterdam, Netherlands) who, together with Alberto Del Bimbo (University of Florence, Italy), provided an initial outline for how such a contest might be conducted. Yining Deng, Ullas Gargi, and Lucy Cherkasova (all of Hewlett-Packard Laboratories, USA) together with Xu Yin (Tsinghua University, China) participated in early discussions based on the original contest outline. Yining Deng and Ullas Gargi implemented a test image collection together with a browser interface. This eased the process for one of us (NJG) to write a first prototype BIRDS-I benchmark in Perl. Ullas Gargi staged the prototype Benchathlon web server at HP Labs. Corinne Jörgensen (SUNY Buffalo, USA) and Alberto Del

Bimbo provided some of the images for the test collection. B.S. Manjunath (U.C. Santa Barbara, USA) provided us with a draft of his forthcoming paper and an explanation of the MPEG–7 normalized rank retrieval rate. Henry Sang, Jr. (Hewlett-Packard Laboratories, USA) provided the server for use at the first Benchathlon contest. Annabelle Eseo (Hewlett-Packard Laboratories, USA) built the secure server that was accessible on the open Internet from behind the firewalls at HP Laboratories.

References

15. TREC (Text REtrieval Conference), benchmark specifications and results http://trec.nist.gov.

16. P.E. Renaud, *Introduction to Client/Server Systems: A Practical Guide for Systems Professionals*, Wiley, 1993.

17. N.J. Gunther, "Images Over the Internet: The Performance Picture," Short Course SC082. SPIE Conference on Internet Imaging, San Jose, CA, Jan. 2000.

18. N.J. Gunther, "On the Application of Barycentric Coordinates to the Prompt and Visually Efficient Display of Multiprocessor Performance Data," *Proc. TOOLS '92*, Edinburgh, Scotland, September 1992.

19. M.H. Kim, "A Geometric Model of Information Retrieval Systems," preprint, Dec. 1999.

20. T. Barclay, J. Gray, D. Slutz, "Microsoft TerraServer: A Spatial Data Warehouse," Technical Report MS–TR–99–29.

21. R.D. Hersch et al., "The Visible Human Slice Web Server: A First Assessment," pp. 253–258, *Proc. SPIE Conference on Internet Imaging, San Jose*, Vol. 3964, CA, Jan. 2000.

22. A. Strupp-Adams and E. Henderson., "Retrieving High Resolution Images Over the Internet from an Anatomical Database," pp. 259–265, *Proc. SPIE Conference on Internet Imaging, San Jose*, Vol. 3964, CA, Jan. 2000.

23. Image Databases. A comprehensive list of image databases available on the Web can be found at http://www-vision.ucsd.edu/~ssantini/dbliterature.html#imgdbases.

24. C. Jörgensen, "Classifying Images: Criteria for Grouping as Revealed in a Sorting Task," *Proc. 6th ASIS SIG/CR Classification Research Workshop*, pp. 65–78, Chicago, Oct. 1995.

25. K. Mahalingam and M.H. Huhns, "A Tool for Organizing Web Information," *IEEE Computer*, pp. 80–83, June 1997.

26. Further background on the Ontology approach is available at http://www.cs.utexas.edu/users/mfkb/related.html.

27. Annotated ground truth database, Department of Computer Science and Engineering, University of Washington, http://www.cs.washington.edu/research/imagedatabase/groundtruth/.

28. D. McG. Squire, H. Müller, and W. Müller, "Improving Response Time by Search Pruning in a Content-Based Image Retrieval System, Using Inverted File Techniques," *IEEE Workshop on Content-based Access of Image and Video Libraries (CBAIVL '99)*, Fort Collins, CO, 1999.

29. R. Buchanan, *The Art of Testing Network Systems*, Wiley, 1996.

30. N.J. Gunther, "A Simple Capacity Model for Massively Parallel Transaction Systems," pp. 1035–1044. *Proc. CMG'93 Conference*, San Diego, CA, Dec. 5–10, 1993.

31. P.S. Salembier and B.S. Manjunath, "Audiovisual Content Description and Retrieval: Tools and MPEG–7 Standardization Techniques," IEEE ICIP '2000, p. 29, 2000.

## Author's Addresses

**Neil J. Gunther.** Performance Dynamics Consulting, 4061 E. Castro Valley Blvd., Suite 110, Castro Valley, CA 94552. njgunther@perfdynamics.com, http://www.perfdynamics.com/.

**Giordano Beretta.** Hewlett-Packard Company, 1501 Page Mill Road, MS 4U-6, Palo Alto, CA 94304. beretta@hpl.hp.com, http://www.hpl.hp.com/personal/Giordano_Beretta/.